\documentclass[12pt,preprint]{aastex}
\usepackage{lscape}

\def\sax{{\sl BeppoSAX }}
\def\ros{{\sl ROSAT }}
\def\gin{{\sl GINGA }}

\def\exo{{\sl EXOSAT }}

\def\xte{{\sl RXTE }}
\def\uhu{{\sl Uhuru }}
\def\swift{{\sl SWIFT }}

\def\xmm{{\sl XMM-Newton }}
\def\chandra{{\sl Chandra }}

\def\cirx1{Cir~X-1~}
\def\b0614{4U~0614+091~}
\def\u1626{4U~1626-67~}
\def\ergsec{\hbox{erg s$^{-1}$ }}
\def\ergcm{\hbox{erg cm$^{-2}$ s$^{-1}$ }}

\def\kms{\hbox{km s$^{-1}$}}
\def\oviii{O~{\sc viii}}

\def\oviiia{O~{\sc viii} L$_{\alpha}$}
\def\oviiib{O~{\sc viii} L$_{\beta}$}
\def\oviib{O~{\sc vii} L$_{\beta}$}

\def\neix{Ne~{\sc ix}}
\def\nex{Ne~{\sc x}}

\def\neii{Ne~{\sc ii}}
\def\neiii{Ne~{\sc iii}}

\def\lapp{\ifmmode\stackrel{<}{_{\sim}}\else$\stackrel{<}{_{\sim}}$\fi}

\def\gapp{\ifmmode\stackrel{>}{_{\sim}}\else$\stackrel{>}{_{\sim}}$\fi}
\def\spose#1{\hbox to 0pt{#1\hss}}

\def\approxlt{\mathrel{\spose{\lower 3pt\hbox{$\sim$}}
        \raise 2.0pt\hbox{$<$}}}
\def\approxgt{\mathrel{\spose{\lower 3pt\hbox{$\sim$}}
        \raise 2.0pt\hbox{$>$}}}

\slugcomment{Submitted for Publication to The Astrophysical Journal}

\shorttitle{SPECTRAL DYNAMICS IN \b0614}

\shortauthors{SCHULZ et al.}

\begin{document}

\title{Dynamical Ne K Edge and Line Variations in the X-Ray Spectrum of the Ultra-compact Binary \b0614}
\author{
N. S. Schulz,\altaffilmark{1}
M. P. Nowak,\altaffilmark{1}
D. Chakrabarty,\altaffilmark{1}
and
C. R. Canizares\altaffilmark{1}
\altaffiltext{1}{Center for Space Research, Massachusetts Institute of Technology,
Cambridge, MA 02139.}}

\begin{abstract}
We observed the ultra-compact binary candidate 4U 0614+091 for a total of 
200 ksec with the high-energy transmission gratings onboard the \chandra
X-ray Observatory. The source is found at various intensity levels with spectral
variations present. X-ray luminosities vary between 2.0$\times10^{36}$ \ergsec
and 3.5$\times10^{36}$ \ergsec. Continuum variations are present at all times and 
spectra can be well fit with a powerlaw component, 
a high kT blackbody component, and a broad line component near oxygen. 
The spectra require adjustments to the Ne K edge and in some occasions
also to the Mg K edge. The Ne K edge appears variable in terms of optical depths and
morphology. The edge
reveals average blue- and red-shifted values implying Doppler velocities of the order
of 3500 \kms. The data show that Ne K exhibits excess column densities of up to several 10$^{18}$
cm$^{-2}$. The variability proves that the excess is intrinsic to the source. The 
correponding disk velocities also imply an outer disk radius of the order of 
$< 10^9$ cm consistent with an ultra-compact binary nature. We also detect a prominent
soft emission line complex near the \oviii L$\alpha$ position which appears extremely broad
and relativistic effects from near the innermost disk
have to be included. Gravitationally broadened line fits also provide nearly edge-on
angles of inclination between 86 and 89$^{\circ}$. 
The emissions appear consistent with an ionized 
disk with ionization parameters of the order of 10$^4$ at radii of a few 10$^7$ cm.
The line wavelengths with respect to \oviiia\ are found variably blue-shifted  
indicating more complex inner disk dynamics.
\end{abstract}


\section{Introduction}

The low mass X-ray binary (LMXB) \b0614 has been persistently bright since
its detection with \uhu~\citep{forman1978}. It has been identified as
a type I X-ray burster~\citep{swank1978} and thus is an accreting 
neutron star. A recent study of its very bright X-ray bursts
lead to a distance estimate of 3.2 kpc~\citep{kuulkers2009} putting
the source a bit beyond earlier projections~\citep{brandt1992}.
It is now confirmed to be an atoll source~\citep{mendez1997} based
on its spectral properties~\citep{singh1994,vanstraaten2000} and timing pattern
~\citep{ford1997}. A peculiar trait of its X-ray spectrum is the 
presence of a hard X-ray tail reaching energies beyond 100 keV
\citep{ford1997} which was modeled as thermal Comptonization
at high electron temperatures of greater than 220 keV
~\citep{piraino1999, fiocchi2008}.
  
The source has recently become a focus of renewed interest for 
several reasons. A very recent survey observation covering the
source from radio, to infrared, optical, UV and X-ray emissions
identified optically thick synchrotron emission from a jet with
radiative powers well beyond $10^{32}$ \ergsec~\citep{migliari2010}.
The jet was found to be present in the source's hard state~\citep{vanstraaten2000}
comparable to what is now found in stellar mass black bole accretors and
active galactic nuclei.

This is even more interesting as many indications place
\b0614 into a rare class of ultra-compact binaries with orbital
periods P $\approxlt 80$ min considered to be near the minimum period
for LMXBs with hydrogen-rich main-sequence donors. \citet{shahbaz2008}
favor a 51.3 min binary period which would further classify this system
as a neutron star accreting either from a hydrogen-deficient star or
a degenerate, likely white dwarf companion~\citep{verbunt1995}. 
Observations with \chandra determined Ne K shell absorption significantly above
the expectation from ISM column densities~\citep{paerels2001}. This 
seems to support the suggestion by~\citet{juett2001} and ~\citet{juett2005} that 
the enhanced Ne/O ratios determined from X-ray spectra point to a low-mass,
neon-rich degenerate white dwarf companion in \b0614 and other
very low period binaries and that these binaries are all ultra-compact.
VLT spectroscopy confirms the deficiency of H and He in the accretion
disk of the proto-type ultra-compact source \u1626 and in \b0614, again
pointing towards an ultra-compact nature~\citep{werner2006}. However, these
data show a peculiar lack of Ne line which seems in stark contrast to the
X-ray findings and it was suggested that the donor could be an eroded C/O
white dwarf with no excessive Ne overabundance.

We observed \b0614 with the high resolution transmission grating spectrometer (HETGS)
onboard \chandra for about 2.5 days in order to study in depth the X-ray absorption
properties. In this paper we specifically focus on the intrinsic source
properties, i.e. the nature of its X-ray continuum emission, possible broad line emissions, 
excess K edge properties, and spectral variability.  

\section{Chandra Observations}

The \chandra HETGS observations were performed during Cycle 10 of the mission and are
part of our guaranteed time program. The detailed observation dates and parameters 
are listed in Table 1. The total amount of 200 ksec of observating time was
arbitrarily split into four pieces of about 40 to 60 ksec duration, 
which were taken over a total time span of
about 4.5 days.
 
\begin{figure}
\includegraphics[angle=0,width=8.5cm]{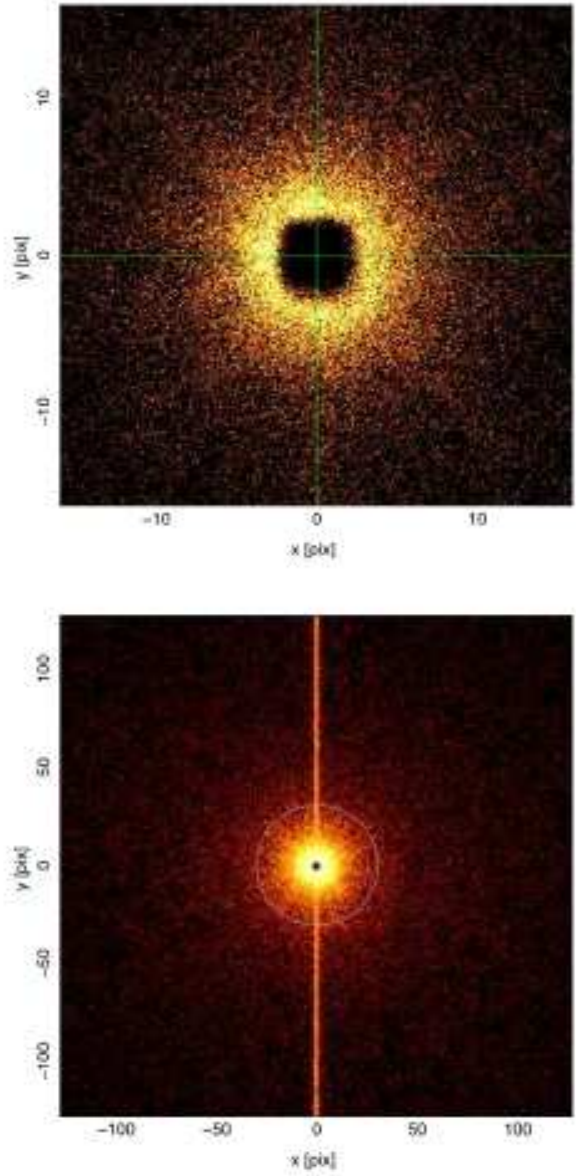}
\caption{The zero oder image of OBSID 10759. The cross in the top
image shows the exact zero order position. In the bottom its exact
coincidence with the readout streak. 
\label{zerorder}}
\end{figure}

All observations were processed using CIAO4.2 with the most recent
CALDB products using the tools offered by the on-line transmission grating catalog
(TGCAT\footnote{see \url{http://tgcat.mit.edu/}}). The zero-order 
point spread function (psf) was sufficiently piled up; therefore an improved 
zero-order postion was determined using \emph{findzo.sl} which uses the 
the intersection of the psf readout-out streak and the HETG dispersion tracks
\footnote{see also \url{http://asc.harvard.edu/ciao/threads/}}.
Figure~\ref{zerorder} shows the zero order point determination for the 
psf in OBSID 10759. Proper positioning is absolutely crucial for this study.
For all the observations we
generated spectra and analysis products for the 1st orders only. 

Pre-screening of the data
revealed variability between observation segments
and given the brightness of the source we integrated one spectrum
for each segment. For the Ne K edge study we also split the
observations into consecutive sub-segments in order to investigate spectral 
changes on smaller time scales. We fitted all spectral orders simultanously, for
plotting purposes we co-added and also re-binned all 1st orders. The spectral
analysis was performed using the latest version 
\emph{ISIS}\footnote{see \url{http://space.mit.edu/ASC/ISIS}} with imported
\emph{Xspec.v12} functions for spectral modeling. Uncertainties ar listed as 90$\%$
confidence limits calculated using the multi-parameter grid search utility
\emph{conf$\_$loop} in \emph{ISIS}.

\subsection{Light Curves and Fluxes
\label{cfluxes}}

\begin{figure}
\includegraphics[angle=0,width=8.5cm]{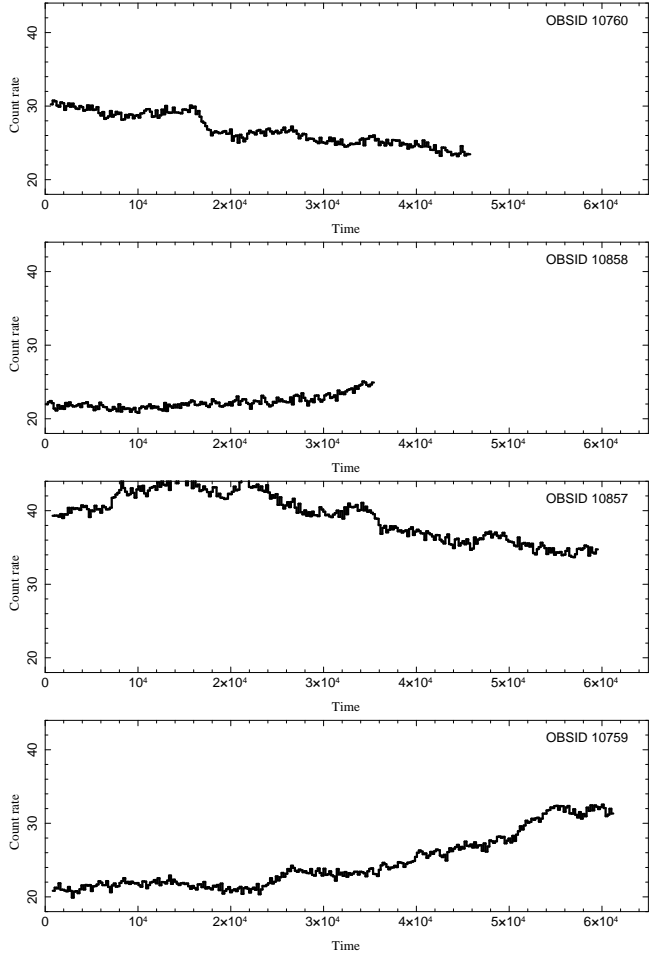}
\caption{The light curves of the four observations. Data bins of 200 sec
contain all HETG 1st order photons from 1.6 to 25~\AA.
\label{lightcurves}}
\end{figure}

Figure~\ref{lightcurves} shows the light curves of the four observations.
The curves, binned to 200 sec bins, appear fairly smooth with mostly gradual
intensity changes. The first three observations are at very similar
intensity levels. The observations show gradual intensity variations at  
a 15$\%$ to 45$\%$ level. Obsid 10857 is brightest
with a factor $\sim$2 higher average intensity. 
There are no rapid changes in the light curves and we also do not detect type I burst
activity. Three observations show intermittent broad flare-like events lasting
for several 1000 sec. 
  
The source fluxes (0.5 -- 10 keV) range from (1.17$\pm0.02)\times10^{-9}$ \ergcm
in OBSID 10858 to (2.05$\pm0.02)\times10^{-9}$ \ergcm in OBSID 10760.
With an interstellar column density of 
3.3$\times10^{21}$ cm$^{-2}$~\citep{piraino1999, migliari2010}
this translates to source luminosities in the range of $(2.0 - 3.5)\times10^{36}$ \ergsec
at a source distance of 3.2 kpc~\citep{kuulkers2009},
which is consistent with a recent analysis of atoll sources by ~\citet{linares2008}
and correponds to about 0.01 L$_{Edd}$.
 
\begin{figure*}
\includegraphics[angle=0,width=16cm]{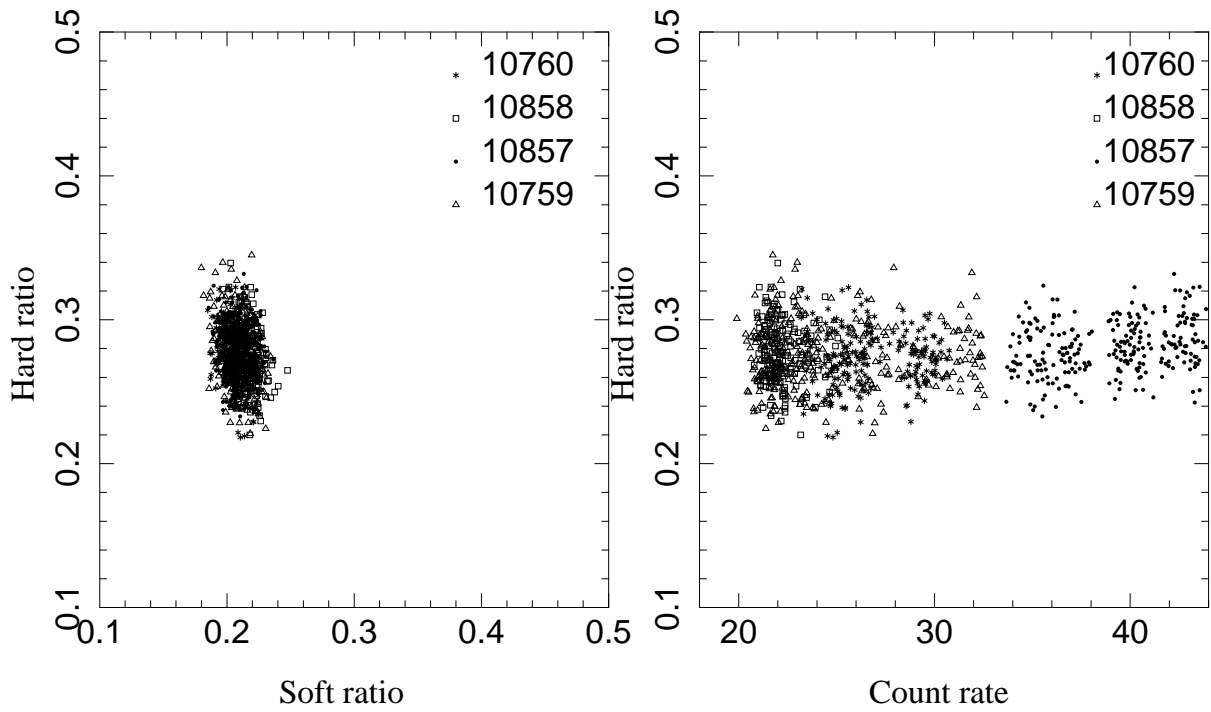}
\caption{{\bf Left:} The HETG 1st order color-color diagram for the four observations. Each
data point represents 200 sec of integration time. Corresponding error bars amount to about
$\pm 0.02$ on each scale. {\bf Right} The corresponding hardness-intensity diagram.
\label{color}}
\end{figure*}

In order to further investigate source state properties we also computed color-color and 
hardness-intensity diagrams, The \chandra HETG bandpass with respect to \xte is limited 
and we do not have the same energy bands to compute hardness ratios 
available as is traditionally done
with broadband instruments such as \xte, \gin, and \exo. As we have done in
\citet{schulz2009}, we chose f$_{xs}$=flux(0.5 keV -- 2.5 keV),
f$_{xm}$=flux(2.5 keV -- 4.5 keV), and f$_{xh}$=flux(4.5 -- 8.0 keV) and thus
for the hard ratio f$_{xh}$/f$_{xm}$ and the soft ratio f$_{xm}$/f$_{xs}$.
The diagram in Figure~\ref{color} (left) shows that the broadband colors of
each observation overlap in a small area. There is no variation in soft ratio,
but some variations in the hard ratio
are similar to the findings by by ~\citet{linares2008} using \xte data. 
The hardness-intensity relations indicate that
that the hard ratio variations are present at all intensities. 
  
\subsection{Continuum Spectra
\label{cspectra}}

Two spectral continuum models have been suggested previously. 
One was provided by ~\citet{piraino1999} which for our relevant 
bandpass suggests a powerlaw plus blackbody model. The other one
was suggested by ~\citet{migliari2010} and consists of a powerlaw
plus a blackbody plus a disk blackbody model. Both models also
featured a soft gaussian line feature in the oxygen region (see also ~\citealt{schulz1999}). 

In all our fits we applied the recently released \emph{Tbnew}
function in \emph{Xspec} which is based on recent high resolution
studies~\citep{juett2004, juett2006} and uses ISM abundance from ~\citet{wilms2000}. 
In order to find an estimate for the interstellar column towards \b0614 
we at first left the NH parameter free without any adjustment to any
elements. The fit at the O K edge provided a column of
$(3.31\pm0.04)\times10^{21}$ cm$^{-2}$ which is consistent with the findings
by previous studies. We then fixed the NH column in the ISM to this value during all fits.  
The powerlaw plus blackbody model alone generally fits 
to our data $<$10~\AA\ with some waviness left in the residuals. 
A hard blackbody as suggested
by both approaches was necessary to fit the spectral band 
$<$3~\AA\ and we add this component into all our fits with
a normalization (A$_{bb} = \sim 1\times10^{35}$ \ergsec at 3.2 kpc)
and temperature ($\sim$ 1.22 keV) very similar to the one suggested
by the studies of ~\citet{piraino1999} and ~\citet{migliari2010}.

\begin{figure*}
\includegraphics[angle=-90,width=16cm]{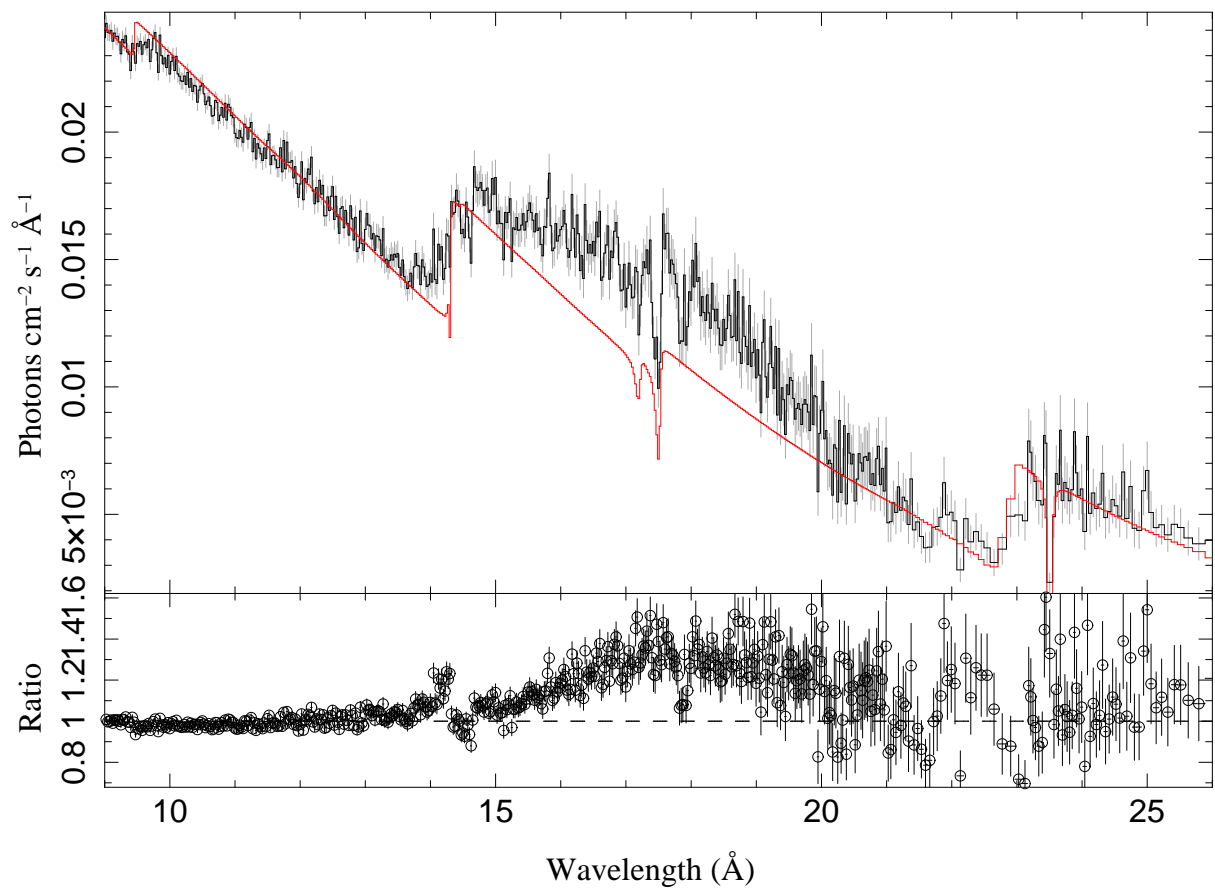}
\caption{The soft X-ray spectrum of obsids 10759 and 10760. The continuum 
is a powerlaw (see Table 2) with an optical depth adjustment of the Ne K 
edge. For the ISM absorption we used the most recent \emph{Tbnew} function
in \emph{XSPEC} which includes the substructure at the Ne and O K edges and
the Fe L edge. The fit also includes a broad Gaussian line near 18\AA. For
illustration purpose we removed the line contribution from the model and
residals. The Ne K also leaves a residual since we do not include any 
wavelength adjustments.
\label{spectrum}}
\end{figure*}

\begin{figure*}
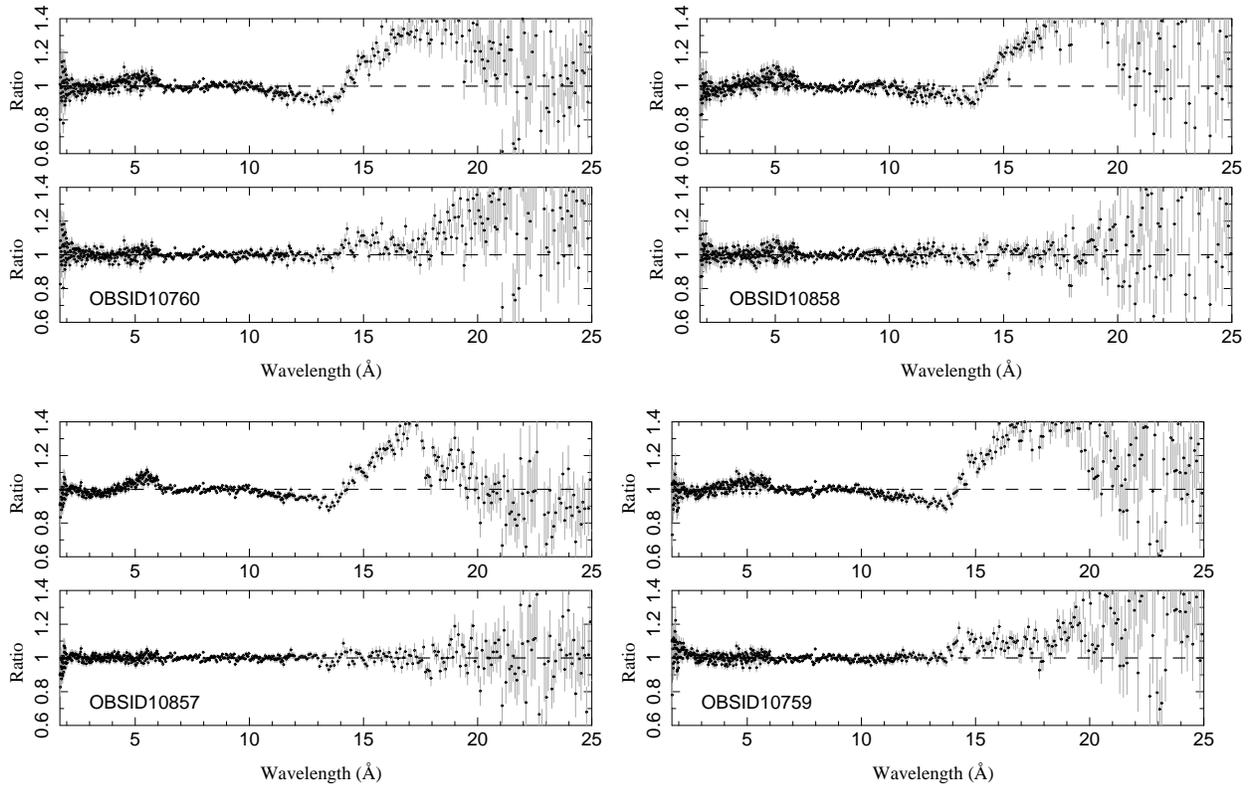

\includegraphics[angle=0,width=8cm]{fig5a.ps}
\includegraphics[angle=0,width=8cm]{fig5b.ps}
\includegraphics[angle=0,width=8cm]{fig5c.ps}
\includegraphics[angle=0,width=8cm]{fig5d.ps}
\caption{Residuals from continuum fits with and without
additional K edge depths and broad line emission. A large impact on the
fits comes from an adjustment of the Ne K edge (14.29~\AA) depth and
the addition of a very broad Gaussian line function around 18~\AA. Some
minor impact arises from an adjustment of the Mg K depth and another
much weaker broad line between 11 and 12~\AA.
\label{residuals}}
\end{figure*}

At higher wavelengths ($>$ 10~\AA) the simple powerlaw approach breaks down and we observe 
significant deviations. Adding a spectral component such as another
soft blockbody or a soft disk blackbody did not result in acceptable fits.
Figure~\ref{spectrum} shows all the relevant features in the soft part
of the spectrum, which consist of O, Ne, and Mg K edges at 22.9, 14.28, and 9.48~\AA,
respectively, an Fe L edge around 17.5~\AA\ and a broad excess around 18~\AA. 
Figure~\ref{residuals}~(top panels without obsid labels) shows the residuals
of the unmodified continuum fits for each observation sequence. 
One strong residual appears at the 
location of the Ne K edge, another 
one is observed between 16 and 20~\AA. Some smaller residuals appeared 
at Mg K and some waviness around 11~\AA\ and in some observations
near 5~\AA. 
In order to avoid adding separate edge components to the broadband analysis
we used the abundance parameter in the 
the \emph{Tbnew} function to fit additional variable edge depths. Our fits show that 
the emission component is so broad   
that the soft residuals can only be fit by a single soft Gaussian line component
similar to the one found in previous analyses~\citep{piraino1999, migliari2010}
and now very recently in \xmm RGS data~\citet{madej2010}.

The new residuals are plotted in Figure~\ref{residuals} (bottom panels
with obsid labels). 
The adjustments provide an acceptable fit to all spectra with reduced
$\chi^2$ values between 1.3 and 1.7 in the worst case. Powerlaw indices were quite
similar in all observations but varied between 2.13 and 2.25, whereas normalizations
A$_{pl}$ varied between 0.35 and 0.65 ph cm$^{-2}$ s$^{-1}$ \AA$^{-1}$.
Table 2 summarizes the continuum components as well as the fits results for the 
broad line. 
In all cases the fits to the spectra required significant  
increases of the Ne K optical depth.
In OBSID 10858 the fits also suggested a significant adjustment to the Mg K edge. 

\section{Ne K Edges Morphology\label{nekedge}}

\begin{figure}
\includegraphics[angle=0,width=16cm]{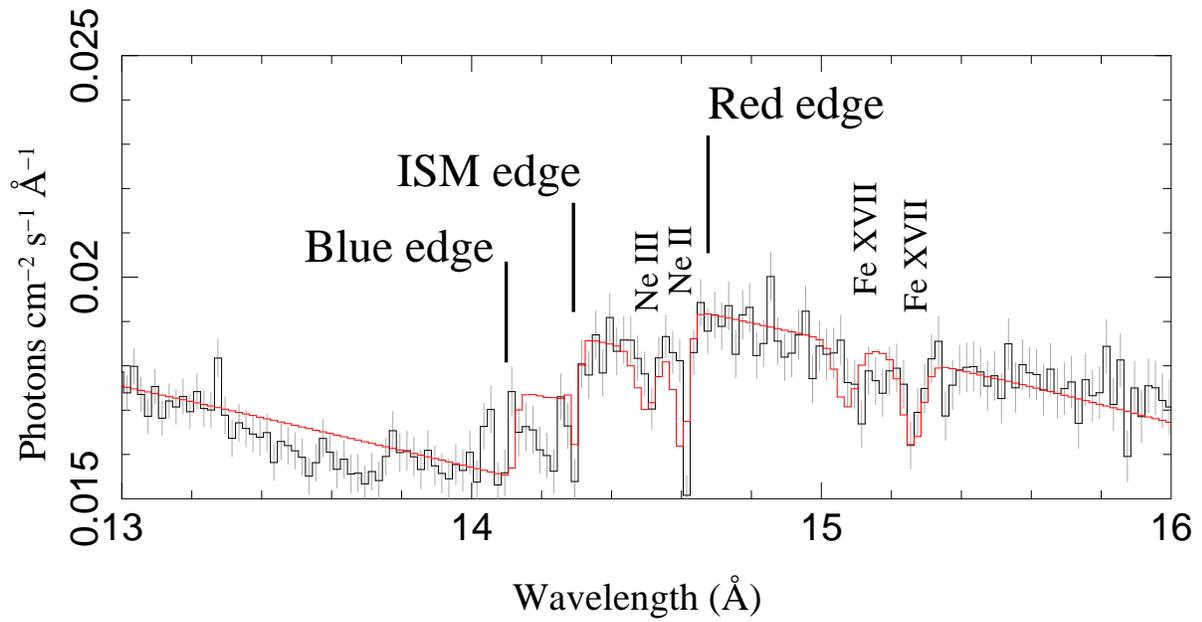}
\caption{Local Ne K edge of the total exposure summed over the
four observations. The edge is very complex showing red- and blue-
shifted components in addition to the ISM contribution. The 
red- and blue-shifted components are smeared as the edge is 
changing at timescales smaller then 1000 sec. The red portion of the 
edge is also perturbed by existing \neii and \neiii lines.
\label{nekexample}}
\end{figure}

In Figure~\ref{nekexample} we show the edge appearance over the entire exposure.
The edge contains red- and blue-shifted components in addition to the ISM contribution.
From the fits it also becomes apparent that the Ne optical
depth changes with time. There is a sharp residual in the fits (see also Figure~\ref{spectrum})
at 14.3~\AA\ which indicates some issues with the actual location of the edge in the spectra.
The interstellar Ne edge structure and value has been
studied in detail by ~\citet{juett2006} and its wavelength has been found to be 14.295$\pm$0.003~\AA\
coincident with the Ne I $1s-3p$ transition consistent with the standard 
 model~\citet{gorczyca2000}.
The following analysis focusses on the local structure and now applies a 
local model consisting of ISM absorption, a powerlaw, and the explicit Ne edge function
in the range between 13 and 15~\AA.
We again fixed the interstellar contribution to the edge to a column density of
3.3$\times10^{21}$ cm$^{-2}$. This density corresponds to an optical depth $\tau_{ISM}$
of 0.17 (N$_{Ne}$ = 4.57$\times10^{17}$ cm$^{-2}$), 
which is more than twice the detection threshold at this particular
wavelength. For the observed continuum levels
an edge with an optical depth of about 0.04 can be detected. Note, that in terms of
significance this only corresponds to about 1$\sigma$ above the continuum. However, in contrast
to line detections which require much higher significances at the centroid wavelength, 
the edge detection threshold is an integral quantity over
several bins above and below the edge (see ~\citealt{schulz2002}). 

We noticed that the morphology of the edge changes on timescales smaller than
our observation segments and we split these segments into several equally timed
and consecutive intervals.
The exposure of these sub-segments was chosen
to provide enough counts in each spectrum for good statistics
during spectral fitting. In conjunction with the good time intervals this lead to
18 sub-segments with 9.7 ksec, one with 8.6 ksec, and one with 5.5 ksec.
We also approximate the edge smearing effect in a very simple way by adding a
blue and a red-shifted component to the Ne edge fits of the sub-segments.  
The optical depths $\tau_{source} = \tau_{bNe}$ + $\tau_{rNe}$
are in excess of the one produced by the interstellar columns,
i.e. $\tau_{Ne}$ = $\tau_{source}$ + $\tau_{ISM}$. 
The fit values vary from sub-segment to sub-segment indicating
the more continuous change with time. This explains the  different appearance of the edge in 
all the segments and sub-segments. Blue-shifted 
depths $\tau_{bNe}$ vary from 0.057 to 0.274, red-shifted depths $\tau_{rNe}$ from 0.013 to 0.245.
The total source depth $\tau_{source}$ (in excess of $\tau_{ISM}$) 
then varies from 0.131$\pm$0.034 to
0.444$\pm$0.130 which corresponds to a Ne column density range of 
3.6$\times10^{17}$ cm$^{-2}$ to 1.2$\times10^{18}$ cm$^{-2}$. 

\begin{figure}
\includegraphics[angle=0,width=16cm]{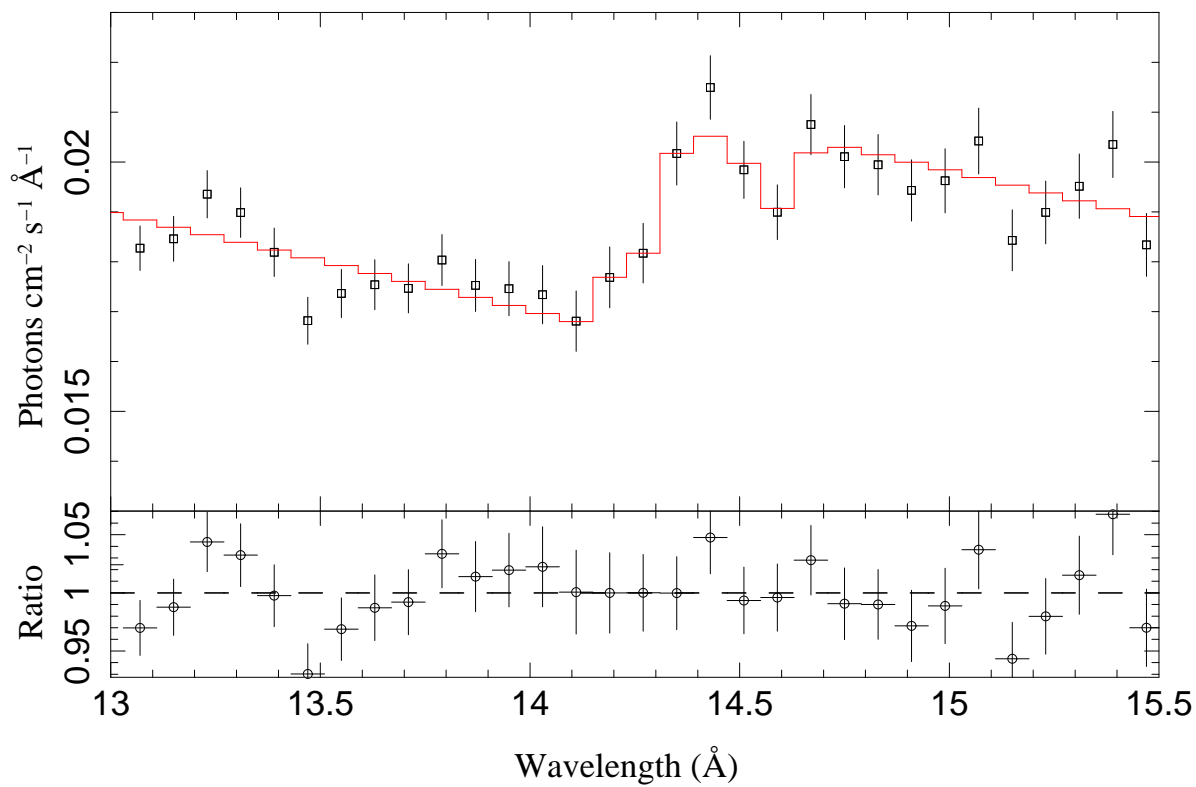}
\includegraphics[angle=0,width=16cm]{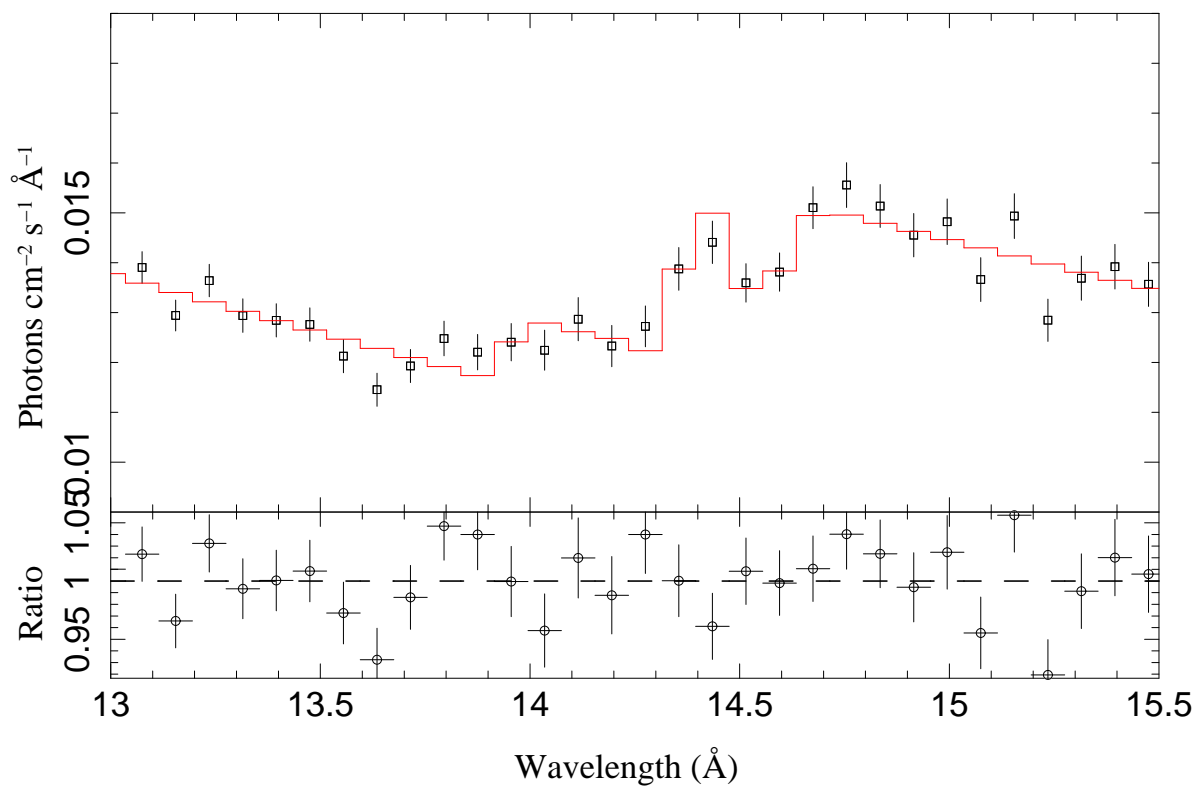}
\caption{Two cases of edge morphology for a very low (top) and 
very large (bottom) velocity smear of the Ne K edge. 
Here we integrated three extreme sub-segments in each case.
Average excess total optical depths vary from 0.10 to 0.14, 
blue- and red-shifts from 0.08 to 0.24 mA.
\label{nek}}
\end{figure}

\begin{figure}
\includegraphics[angle=0,width=16cm]{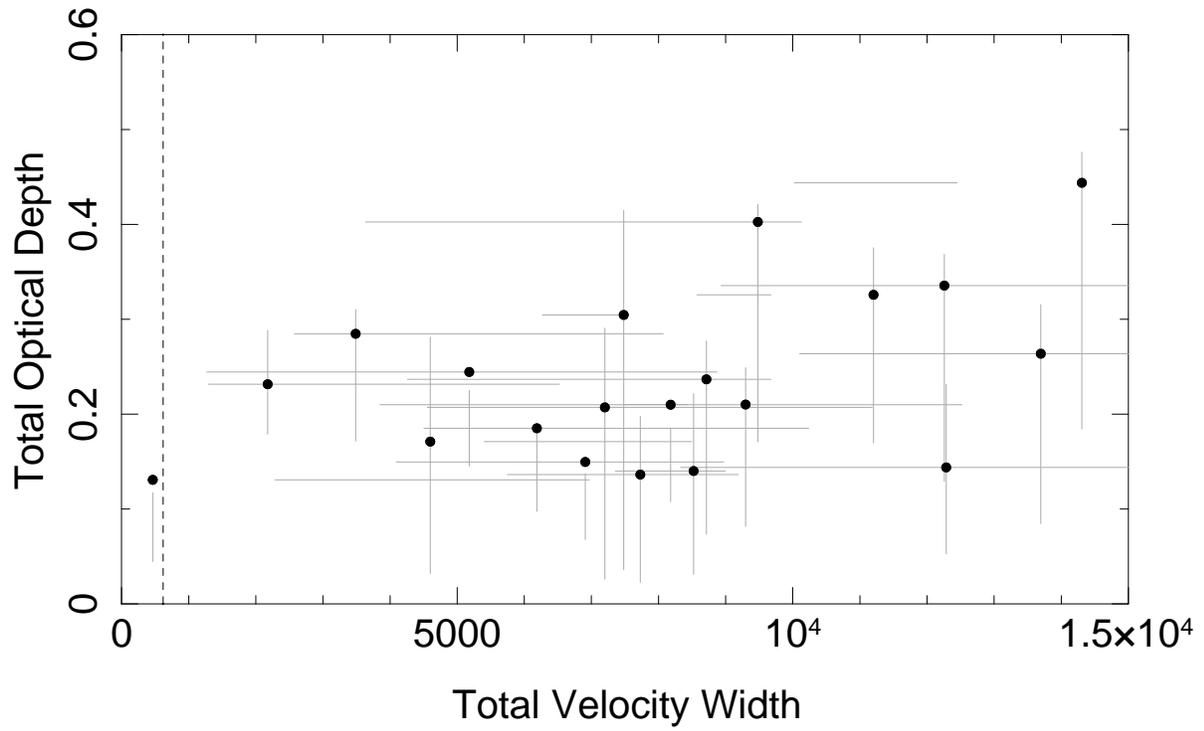}
\caption{The total velocity spread (blue- and red-shifts) plotted against
the total intrinsic edge optical depth. The dotted line marks the
likely amount of a velocity smear from an
orbital period of 51.3 min~\citet{shahbaz2008}.
\label{edgeparams}}
\end{figure}

Figure~\ref{nek}
shows two examples of edge smear, one with a low spread of about 0.15 \AA\ (top), and one
with a high spread of about 0.65 \AA. 
The measured shifts imply velocity smears ranging from about 2000 to 15000 \kms as
plotted in Figure~\ref{edgeparams}. The measured blue-shifts
show variations in all observation segments. 
Figure~\ref{edgeparams} plots the total velocity smear which is comprised of 
added shifts of the edge on the red and blue side versus the similarly added optical 
depths ($\tau_{source}$). Note that observed smears are generally higher than the one expected
from an orbital binary period of 51.3 min~\citep{shahbaz2008}. The uncertainties specifically
for high values are fairly large and most segment values agree with values between
7000 and 8000 \kms, which after subtracting a possible binary orbital contribution translates
to dynamic velocities between 3100 \kms and 3700 \kms.  

The resulting optical depth and smear velocities do not seem to reveal any specific pattern with
time. Nevertheless, we tested if we could find recurring patterns by folding the light curve
by several periods in the system as suggested by ~\citet{shahbaz2008}. We used OBSID 10759
for such a test and generated and fitted six phase-binned spectra. 
We applied periods of 41 min, 51 min, 62 min, and
120 min. In no cases did we observe evolving patterns,
which leads us to conclude that the observed variations
are random with respect to possible short orbital periods. We do not have enough statistics
to fully investigate the time scale of change in the edge morphology other than observe that 
it is likely of the order of $<< 10^4$ sec.   

\subsection{Broad Line Emission\label{emission}}

Previous studies always required to include a strong broad emission line feature near the 
oxygen region into any spectral fit of \b0614~\citep{singh1994, piraino1999, schulz1999,
migliari2010}. The HETG spectra also require such a feature between 17 and 19~\AA\ as 
is shown in the residuals in Figure~\ref{spectrum} and ~\ref{residuals}. 
The feature is broad and is fully resolved in all fits. 
However, the line shows a very high avarage $\sigma$ width of 1.80$\pm$0.13~\AA.

The broad line can be centroided to a wavelength accuracy of almost 1$\%$ and we find
wavelengths which appear more consistent in obsids 10760, 10759, and 10857 at 
an average value of 17.7~\AA\ (0.700 keV), but significantly different in obsid 10858 at
18.2~\AA\ (0.681 keV). This indicates that the feature centroid is variable. Furthermode,
we cannot identify a single responsible ion at these wavelengths. If it is 
a wavelength complex it has to include major lines we expect in this region, 
which are expected to be predominantly from Fe XVII and the O~VII and O~XVIII Lyman series.
However, we simulated several cases of broad line complexes using the embedded
XSTAR function \emph{photemis}, which is available in the most recent \emph{Xspec}
version as a local module, and find that Fe XVII cannot contribute because
we do not observe a significant contribution of the most prominent line at 15~\AA. 
The most likely solution is the O XVIII Lyman series with a dominant contribution
from the O XVIII Lyman $\alpha$ line. The rest wavelength of this
line is 18.97~\AA\ (0.656 eV) and the measured centroid position then 
appears significant blue-shifted between 0.77~\AA\ and 1.27~\AA. 

An identification with a single \oviiia line leads to velocity broadening
of almost 3.0$\times10^4$ \kms, which has to be considered relativistic.
As done for the \xmm RGS data~\citep{madej2010} we fit 
this line feature with a gravitationally broadened line model 
\emph{laor}~\citep{laor1991} available in \emph{Xspec}. 
The fits are summarized in Table 3. The model contributes to the fits at least as well 
as the broad gaussian with some improvement in the soft residuals. For 1540 d.o.f. we
obtain reduced $\chi_{nu}^2$ between 1.37 and 1.63 in the band between 1.7 and 25\AA.
The \emph{laor} fits produce two fit solutions with respect to inclination which
appear equivalent. In obsids 10857 and 10759 the fit finds a value around
86$^{\circ}$ and r$_{in}$ of $\sim$9, in obsids 10760 and 10858 values of
89$^{\circ}$ and r$_{in}$ of $\sim$2 are preferred. In Table 3 we fixed the 
inclination to 86$^{\circ}$. The fits also show line wavelengths 
blue-shifted with respect to the \oviiia line rest wavelength by amounts similar
to the one observed in the gaussian fits of up to 6$\%$. While the \emph{laor} model
predicts a blue-shift due to gravitiational effects, shifts in Table 3 still
appear significantly variable up to 70$\%$.


\section{Discussion}

Our analysis shows that \b0614 is observed in a hard state. The continuum is
dominated by a powerlaw of index of about 2.2 similar to the one suggested
by ~\citet{piraino1999} and very recently by ~\citet{madej2010}. 
We also confirm the need for a hard blackbody of
about 1.2 keV as suggested by ~\citet{piraino1999} and ~\citet{migliari2010}.
However, the spectra do not show a dominant disk blackbody component as 
found by ~\citet{migliari2010} in \swift data. On the other hand, we searched
for an Fe K edge as suggested by ~\citet{piraino1999} indicating the existence
of a highly absorbed high energy component to no avail.

\b0614 has been classified as an ultra-compact binary system. Specifically the 
detection of an excessive Ne K edge in an early \chandra 
observation~\citep{paerels2001} sparked
interpretations that in combination with a possible ultra-short orbital period 
the donor stars are C-O or O-Ne-Mg stars~\citep{juett2001}. Our observations
indeed confirm the existence excess Ne K edge optical depth. 
A recent VLA study on the other hand finds a rather different
account and observed Ne~II and~III emissions which suggested Ne to be underabundant
~\citep{werner2006}. 
Our new observations show that the Ne K edge clearly shows
an excess source intrinsic component. The edge is highly variable which is evidence
that the excess Ne is source intrinsic. We conclude on this basis that \b0614
indeed shows a Ne excess abundance. Whether this overabundance is restricted
to Ne alone is uncertain. In at least one observation we find excess Mg as well.
For other elements we do not find any conclusive overabundance signatures.

Another aspect of our study is the dynamic appearence of the edge.
We find a smeared edge which we approximate by blue- and redhsifted components. To our 
knowledge this is the first time this kind of dynamics
has been observed in a cool K edge. 
We tested this dynamics with known periodicities~\citep{shahbaz2008}
and conclude that it is random in nature. The Ne edge shows systematic
blue- and red-shifts at average Doppler velocities of around 3500 \kms.
If we associate this cool absorbing material with an extended outer edge rim of the 
accretion disk and near edge-on view 
we can infer an upper limit to the outer radius of the disk of less than 
10$^9$ cm (or $<$3500 \kms) consistent with an ultra-compact binary. In order to
make this feasibkle we have to assume that the disk radius is small enough in order to
have X-ray with blue- and red-shifted absorption scattered in our line of sight. 
This may not be unfeasible but needs future detailed modeling. The random appearance
of optical depth variations as well as edge shifts may also signal the presence of
heavy turbulence of cool plasma entering the accretion resulting in heavy convulsions
producing red- and and blue-shifts at times.

There is at least one broad line emission line complex 
in the spectra near ionized oxygen, which has already been
seen in several previous observations with \exo~\citep{singh1994}, \ros~\citep{schulz1999},
\sax~\citep{piraino1999}, and \swift~\citep{migliari2010}. In neither of these obervations
could the line be resolved because of insufficient instrument resolution.
Even though the HETG in our study fully resolves the line, it still finds 
it broad on previously suggested levels and does
not show structure. We find the width between 50 and 66 eV
similar to previous studies. Very recent \xmm observations also find a single broad line in
RGS spectra \citep{madej2010} and it is quite obvious that the line feature cannot
be further resolved. 

The line location suggests that it is dominated by \oviiia\
with maybe some contribution of \oviib~ and \oviiib~ emission. A single gaussian line interpretation  
leads to Doppler velocities of the order of 3$\times10^4$ \kms, which is 
10$\%$ of the speed of light and here we need to include relativistic effects
such as gravitational (GR) broadening. 
\citet{madej2010} successfully fit a GR profile to the line providing
a line centroid of 18.5\AA, an inner radius of 3.5 R$_g$, and a line of sight inclination
of 88$^{\circ}$. Our fits to the four observation segments give similar results,
however also produce a larger spread of values.
The inclination angles appears to be similar, an averaged value gives
$87.5^{\circ}$. 

If these broad lines originate from the accretion disk, a viewing
angle near 90$^{\circ}$ makes sense and is already argued for in ~\citet{migliari2010}.
In general, disk velocities of 3$\times10^4$ \kms~
already require disk radii of a few $10^7$ cm and significantly lower inclination
values would push the emission site beyond the inner edge of the disk. 
With a source luminosity of 0.02 L$_{edd}$ this 
requires ionization parameters far beyond $10^4$ and consequently barely allow for ionization fractions
for H-like ions. We have observed high ionization parameters in the high flux state
of Cir X-1, where edge-on viewing also seems likely~\citep{brandt2000, schulz2008}.
However, there the higher luminosities produce measurable ionization fractions
at much larger disk radii.
In the case of \b0614 lower inclinations thus would lead to unrealistic dynamics, 
emission radii, and ionization parameters. 
We see some resemblance to the case of \u1626~\citep{schulz2001, krauss2007}, which being viewed
face on already revealed disk (v sin i) velocities of several thousand \kms. 
\u1626 viewed edge-on would also likely show high velosity broadening in its lines.
Recently the presence of soft broad line emission has also been reported
in the Z-sources GX 349+2~\citep{iaria2009} and Cyg X-2~\citep{schulz2009}.

An identification of \oviiia would also imply blue-shifts larger than 0.5~\AA\ independent
of the line model. Even though the gravitational model predicts some blue-shift~\citep{laor1991},
it should be consistent in all observations. However, the observed line wavelengths
show significant variations indicating some additional dynamics in the line emitting
region. Furthermore, \chandra lines show consistently higher blus-shifts compared
to the \xmm measurement~\citep{madej2010}. 
Possibilities may include changes in ionization balance, variable Compton scattering,
or inner disk outflows. Changes in ionization balance are not obvious as observed spectral
and flux changes do not appear dramatic. Compton upscattering requires very high 
optical depths as can be found in the disk but should lead to some tail on the blue side 
of the line which is not obvious in our data. An outflow interpretation remains viable, 
but needs more detailed modeling in order to quantify further. The difference in shift is
about 0.7~\AA\ corresponding to an order of $10^4$ \kms. 
Observed outflow velocities of this amount are rare in Galactic binaries, though. 
Winds and outflows in X-ray binaries 
are now commonly observed, mostly in systems exhibiting strong
X-ray luminosities, such as in \cirx1~\citep{brandt2000, schulz2002a},
GX 13+1~\citep{ueda2004} or
the blackhole binary GRS J1655-40~\citep{miller2006}, but all at much lower velocities. 
However, strong
outflows of several 1000 \kms\ now also have been observed in sub-Eddington accretors such
as 4U~1822-37~\citep{bayless2010} and Her X-1~\citep{boroson2007} in UV spectra. \b0614
has gained a reputation in recent years of being of specifically violent
nature showing sporadic abnormal type I superbursts~\citep{kuulkers2009} and
an active jet during its hard state quite similar to what is commonly
observed in black hole binaries~\citep{migliari2010}. 

There are also several issues with respect to the observed abundances in the spectra.
If \b0614 is indeed ultra-compact we would expect that
the donor star is either a C-O or O-Ne or similar white dwarf. In this respect it
makes sense that we observe prodominantly neon and oxygen-rich matter in the accretion
disk. However, some of the findings raise more questions. First of all, if we indeed
see highly ionized oxygen in the inner disk, why are there no similar \neix\ and \nex\ emissions.
The emissivity distributions with respect to temperature of these ions strongly overlap and, like
in the case of \u1626, we whould observe similar strong emissions from O and Ne. Our 
spectra allow for broad line flux upper limits at \neix\ and \nex\ wavelengths which are 
over an order of magnitude fainter than what we observe in \oviiia\ which does not provide
an answer to the problem. Likewise, our analysis of the K edges reveals significant excesses
in the Ne K columns, but not in the O K columns. At radii of the order of $10^9$ cm 
and the observed source luminosities we do not expect ionization parameters high enough to
destroy all neutral oxygen.

Even though we lack answers to many of our findings, we tentatively conclude the following:
\begin{itemize}
\item{\b0614 shows a variable excess Ne absorption column densities
of up to several 10$^{18}$ cm$^{-2}$}
\item{The overabundance has a dynamic high velocity component and is evidently source intrinsic}
\item{The Ne K edge velocity smear could point to a probable outer disk radius limit of 
several $10^9$ cm supporting an ultra-compact binary nature}
\item{The soft X-ray line emission appears extremely broad with corresponding 
velocities of 3$\times10^4$ \kms and likely originates from a relativistic disk regime.}
\item{The line emission indicates an ionized disk layer, predominantly \oviiia,
at radii well below $10^8$ cm
down to only several gravitational radii.}
\item{The soft X-ray line emission shows variable blue-shifts with respect to the \oviiia rest
wavelength indicating further dynamical processes near the inner disk.}
\end{itemize}

\acknowledgments
We thank all the members of the \chandra\ team for their enormous efforts,
specifically D. P. Huenemoerder, J. Davis, and J. Houck for easing
data processing and fitting procedures.

\bibliographystyle{jwapjbib}

\begin{thebibliography}{}

\bibitem[\protect\astroncite{{Bayless} et~al.}{2010}]{bayless2010}
{Bayless}, A.~J., {Robinson}, E.~L., {Hynes}, R.~I., {Ashcraft}, T.~A., \&
  {Cornell}, M.~E.,  2010, \apj, 709, 251

\bibitem[\protect\astroncite{{Boroson} et~al.}{2007}]{boroson2007}
{Boroson}, B.~S., {Vrtilek}, S.~D., {Raymond}, J.~C., \& {Still}, M.,  2007,
  \apj, 667, 1087

\bibitem[\protect\astroncite{{Brandt} et~al.}{1992}]{brandt1992}
{Brandt}, S., {Castro-Tirado}, A.~J., {Lund}, N., {Dremin}, V., {Lapshov}, I.,
  \& {Syunyaev}, R.,  1992, \aap, 262, L15+

\bibitem[\protect\astroncite{{Brandt} \& {Schulz}}{2000}]{brandt2000}
{Brandt}, W.~N., \& {Schulz}, N.~S.,  2000, \apjl, 544, L123

\bibitem[\protect\astroncite{{Fiocchi} et~al.}{2008}]{fiocchi2008}
{Fiocchi}, M., {Bazzano}, A., {Ubertini}, P., {Bird}, A.~J., {Natalucci}, L.,
  \& {Sguera}, V.,  2008, \aap, 492, 557

\bibitem[\protect\astroncite{{Ford} et~al.}{1997}]{ford1997}
{Ford}, E.~C., et~al., 1997, \apjl, 486, L47+

\bibitem[\protect\astroncite{{Forman} et~al.}{1978}]{forman1978}
{Forman}, W., {Jones}, C., {Cominsky}, L., {Julien}, P., {Murray}, S.,
  {Peters}, G., {Tananbaum}, H., \& {Giacconi}, R.,  1978, \apjs, 38, 357

\bibitem[\protect\astroncite{{Gorczyca}}{2000}]{gorczyca2000}
{Gorczyca}, T.~W.,  2000, \pra, 61, 024702

\bibitem[\protect\astroncite{{Iaria} et~al.}{2009}]{iaria2009}
{Iaria}, R., {D'A{\'{\i}}}, A., {di Salvo}, T., {Robba}, N.~R., {Riggio}, A.,
  {Papitto}, A., \& {Burderi}, L.,  2009, \aap, 505, 1143

\bibitem[\protect\astroncite{{Juett} \& {Chakrabarty}}{2005}]{juett2005}
{Juett}, A.~M., \& {Chakrabarty}, D.,  2005, \apj, 627, 926

\bibitem[\protect\astroncite{{Juett}, {Psaltis} \&
  {Chakrabarty}}{2001}]{juett2001}
{Juett}, A.~M., {Psaltis}, D., \& {Chakrabarty}, D.,  2001, \apjl, 560, L59

\bibitem[\protect\astroncite{{Juett}, {Schulz} \&
  {Chakrabarty}}{2004}]{juett2004}
{Juett}, A.~M., {Schulz}, N.~S., \& {Chakrabarty}, D.,  2004, \apj, 612, 308

\bibitem[\protect\astroncite{{Juett} et~al.}{2006}]{juett2006}
{Juett}, A.~M., {Schulz}, N.~S., {Chakrabarty}, D., \& {Gorczyca}, T.~W.,
  2006, \apj, 648, 1066

\bibitem[\protect\astroncite{{Krauss} et~al.}{2007}]{krauss2007}
{Krauss}, M.~I., {Schulz}, N.~S., {Chakrabarty}, D., {Juett}, A.~M., \&
  {Cottam}, J.,  2007, \apj, 660, 605

\bibitem[\protect\astroncite{{Kuulkers}, {in't Zand} \&
  {Lasota}}{2009}]{kuulkers2009}
{Kuulkers}, E., {in't Zand}, J.~J.~M., \& {Lasota}, J.,  2009, \aap, 503, 889

\bibitem[\protect\astroncite{{Laor}}{1991}]{laor1991}
{Laor}, A.,  1991, \apj, 376, 90

\bibitem[\protect\astroncite{{Linares}, {van der Klis} \&
  {Wijnands}}{2008}]{linares2008}
{Linares}, M., {van der Klis}, M., \& {Wijnands}, R.,  2008,
\newblock in AAS/High Energy Astrophysics Division, Vol.~10, 10.17

\bibitem[\protect\astroncite{{Madej} et~al.}{2010}]{madej2010}
{Madej}, O., {Jonker}, P., Fabian, A., Pinto, C., Verbunt, F.~W.~M., \&
  de~Plaa, J.,  2010,
  http://www.sron.nl/index.php?option=com{$\_$}content{$\&$}task=view{$\&$}id=%
2665{$\&$}Itemid=2354, 0

\bibitem[\protect\astroncite{{Mendez} et~al.}{1997}]{mendez1997}
{Mendez}, M., {van der Klis}, M., {van Paradijs}, J., {Lewin}, W.~H.~G.,
  {Lamb}, F.~K., {Vaughan}, B.~A., {Kuulkers}, E., \& {Psaltis}, D.,  1997,
  \apjl, 485, L37+

\bibitem[\protect\astroncite{{Migliari} et~al.}{2010}]{migliari2010}
{Migliari}, S., et~al., 2010, \apj, 710, 117

\bibitem[\protect\astroncite{{Miller} et~al.}{2006}]{miller2006}
{Miller}, J.~M., {Raymond}, J., {Fabian}, A., {Steeghs}, D., {Homan}, J.,
  {Reynolds}, C., {van der Klis}, M., \& {Wijnands}, R.,  2006, \nat, 441, 953

\bibitem[\protect\astroncite{{Paerels} et~al.}{2001}]{paerels2001}
{Paerels}, F., et~al., 2001, \apj, 546, 338

\bibitem[\protect\astroncite{{Piraino} et~al.}{1999}]{piraino1999}
{Piraino}, S., {Santangelo}, A., {Ford}, E.~C., \& {Kaaret}, P.,  1999, \aap,
  349, L77

\bibitem[\protect\astroncite{{Schulz}}{1999}]{schulz1999}
{Schulz}, N.~S.,  1999, \apj, 511, 304

\bibitem[\protect\astroncite{{Schulz} \& {Brandt}}{2002}]{schulz2002a}
{Schulz}, N.~S., \& {Brandt}, W.~N.,  2002, \apj

\bibitem[\protect\astroncite{{Schulz} et~al.}{2002}]{schulz2002}
{Schulz}, N.~S., {Canizares}, C.~R., {Lee}, J.~C., \& {Sako}, M.,  2002, \apjl,
  564, L21

\bibitem[\protect\astroncite{{Schulz} et~al.}{2001}]{schulz2001}
{Schulz}, N.~S., {Chakrabarty}, D., {Marshall}, H.~L., {Canizares}, C.~R.,
  {Lee}, J.~C., \& {Houck}, J.,  2001, \apj, 563, 941

\bibitem[\protect\astroncite{{Schulz} et~al.}{2009}]{schulz2009}
{Schulz}, N.~S., {Huenemoerder}, D.~P., {Ji}, L., {Nowak}, M., {Yao}, Y., \&
  {Canizares}, C.~R.,  2009, \apjl, 692, L80

\bibitem[\protect\astroncite{{Schulz} et~al.}{2008}]{schulz2008}
{Schulz}, N.~S., {Kallman}, T.~E., {Galloway}, D.~K., \& {Brandt}, W.~N.,
  2008, \apj, 672, 1091

\bibitem[\protect\astroncite{{Shahbaz} et~al.}{2008}]{shahbaz2008}
{Shahbaz}, T., {Watson}, C.~A., {Zurita}, C., {Villaver}, E., \&
  {Hernandez-Peralta}, H.,  2008, \pasp, 120, 848

\bibitem[\protect\astroncite{{Singh} \& {Apparao}}{1994}]{singh1994}
{Singh}, K.~P., \& {Apparao}, K.~M.~V.,  1994, \apj, 431, 826

\bibitem[\protect\astroncite{{Swank} et~al.}{1978}]{swank1978}
{Swank}, J.~H., {Boldt}, E.~A., {Holt}, S.~S., {Serlemitsos}, P.~J., \&
  {Becker}, R.~H.,  1978, \mnras, 182, 349

\bibitem[\protect\astroncite{{Ueda} et~al.}{2004}]{ueda2004}
{Ueda}, Y., {Murakami}, H., {Yamaoka}, K., {Dotani}, T., \& {Ebisawa}, K.,
  2004, \apj, 609, 325

\bibitem[\protect\astroncite{{van Straaten} et~al.}{2000}]{vanstraaten2000}
{van Straaten}, S., {Ford}, E.~C., {van der Klis}, M., {M{\'e}ndez}, M., \&
  {Kaaret}, P.,  2000, \apj, 540, 1049

\bibitem[\protect\astroncite{{Verbunt} \& {van den Heuvel}}{1995}]{verbunt1995}
{Verbunt}, F., \& {van den Heuvel}, E.~P.~J.,  1995,
\newblock in X-ray binaries, p. 457 - 494, ed. {W.~H.~G.~Lewin, J.~van
  Paradijs, \& E.~P.~J.~van den Heuvel}, 457

\bibitem[\protect\astroncite{{Werner} et~al.}{2006}]{werner2006}
{Werner}, K., {Nagel}, T., {Rauch}, T., {Hammer}, N.~J., \& {Dreizler}, S.,
  2006, \aap, 450, 725

\bibitem[\protect\astroncite{{Wilms}, {Allen} \& {McCray}}{2000}]{wilms2000}
{Wilms}, J., {Allen}, A., \& {McCray}, R.,  2000, \apj, 542, 914

\end{thebibliography}

\begin{table*}
\begin{center}
{\sc TABLE~1: CHANDRA HETGS X-RAY OBSERVATIONS IN 2009}
\begin{tabular}{lcccc}
            & & & & \\
\tableline
 Obsid &  Start Date & Start Time & Exposure & HETG 1st rate\\
       & [UT]        & [UT]       & [ks]     & cts s$^{-1}$ \\
\tableline
            & & & & \\
 10760 & Jan 18 2009 & 23:07:34 &  44.1 & 13.89  \\
 10858 & Jan 19 2009 & 17:45:01 &  34.4 & 12.33  \\
 10857 & Jan 21 2009 & 13:49:21 &  57.3 &  9.90  \\
 10759 & Jan 24 2009 & 05:07:34 &  60.3 & 24.29  \\
            & & & & \\
\tableline
\end{tabular}
\end{center}

\label{tab:obs}
\end{table*}

\begin{landscape}
\begin{table*}
\begin{center}
\vbox{
\large
\center
{\sc TABLE~2 SPECTRAL FIT: POWERLAW + BBODY + GAUSSIAN LINE PARAMETERS}
\vskip 4pt
\begin{tabular}{lccccccc}
\hline
\hline
 OBSID & A$_{pl}$ & $\Gamma$ & A$_{bb}$ & kT$_{bb}$ & A$_{line}$ & $\lambda_{line}$ & $\sigma_{line}$\\
       & ph \AA$^{-1}$~cm$^{-2}$~s$^{-1}$ &  -- & 10$^{36}$ erg s$^{-1}$/D$_{10kpc}^2$ & keV & ph~cm$^{-2}$~s$^{-1}$ & \AA & \AA  \\ 
\hline
 & & & & & & \\
10760 &  0.450$^{  0.006}_{  0.006}$  &  2.244$^{  0.009}_{  0.009}$ &  0.76
$^{  0.06}_{  0.05}$  &  1.360$^{  0.114}_{  0.153}$ & 0.062$^{0.005}_{0.003}$ &
17.752$^{0.176}_{0.177}$ & 1.748$^{0.132}_{0.137}$\\
10858 &  0.368$^{  0.007}_{  0.003}$  &  2.211$^{  0.013}_{  0.008}$ &  1.04
$^{  0.05}_{  0.05}$  &  1.294$^{  0.091}_{  0.026}$ & 0.065$^{0.002}_{0.001}$ &
18.193$^{0.144}_{0.155}$ & 2.066$^{0.167}_{0.133}$\\
10857 &  0.599$^{  0.006}_{ 0.005}$  &  2.134$^{  0.006}_{  0.008}$ &  1.04
$^{  0.05}_{  0.05}$  &  1.389$^{  0.091}_{  0.096}$ & 0.056$^{0.009}_{0.006}$ &
17.534$^{0.118}_{0.134}$ & 1.729$^{0.111}_{0.139}$\\
10759 &  0.415$^{  0.005}_{  0.004}$  &  2.246$^{  0.008}_{  0.007}$ &  1.04
$^{  0.05}_{  0.05}$  &  1.389$^{  0.091}_{  0.096}$ & 0.058$^{0.005}_{0.004}$ &
17.835$^{0.165}_{0.147}$ & 1.755$^{0.117}_{117}$\\
 & & & & & & \\
\hline
\end{tabular}
\normalsize
}
\end{center}

\label{confits}
\end{table*}
\end{landscape}

\begin{table*}
\begin{center}
\vbox{
\large
\center
{\sc TABLE~3 SPECTRAL FIT PARAMETERS OF A RELATIVISTIC LAOR PROFILE}
\vskip 4pt
\begin{tabular}{lccccccc}
\hline
\hline
 OBSID & A$_{l}$ & E$_{l}$ & $\lambda_l$ & r$_{in}$ & Incl. & $\chi_{nu}^2$\\
       & ph \AA$^{-1}$~cm$^{-2}$~s$^{-1}$ & keV & \AA & GM/c$^2$ & deg. &  \\ 
\hline
 & & & & & &\\
10760 &  0.063$^{  0.004}_{  0.003}$  &  0.685$^{  0.001}_{  0.004}$ &  18.10
$^{  0.10}_{  0.03}$  &  12.87$^{0.36}_{2.05}$ & 86 & 1.62\\
10858 &  0.064$^{  0.003}_{  0.007}$  &  0.672$^{  0.002}_{  0.001}$ &  18.44
$^{  0.05}_{  0.04}$  &  10.85$^{2.15}_{1.33}$ & 86 & 1.27\\
10857 &  0.062$^{  0.003}_{  0.004}$  &  0.698$^{  0.001}_{  0.001}$ &  17.76
$^{  0.04}_{  0.02}$  &  9.47$^{1.64}_{0.59}$ & 86.2$\pm$0.5 & 1.37\\
10759 &  0.057$^{  0.005}_{  0.004}$  &  0.694$^{  0.002}_{  0.001}$ &  17.88
$^{  0.03}_{  0.06}$  &  8.02$^{1.04}_{1.01}$ & 86.1$\pm$0.5 & 1.63\\
 & & & & & & \\
\hline
\end{tabular}
\normalsize
}
\end{center}

\label{linefits}
\end{table*}

\end{document}